\documentclass[aps,prb,reprint]{revtex4-1}

\usepackage{graphicx}
\usepackage{enumerate}
\usepackage{hyperref}
\usepackage{textcomp}
\usepackage{amsmath}
\usepackage{amssymb}
\usepackage{pgf}
\usepackage{bm}
\usepackage{epsfig}
\usepackage{figsize}

\begin{document}

\title{Effect of exchange interaction on electronic instabilities in the honeycomb lattice:\\ A functional renormalization group study}

\author{Song-Jin O}
\email{phys5@ryongnamsan.edu.kp}
\author{Yong-Hwan Kim, Ho-Yong Rim, Hak-Chol Pak}
\author{Song-Jin Im}
\email{sj.im@ryongnamsan.edu.kp}
\affiliation{Faculty of Physics, Kim Il Sung University, Taesong District, Pyongyang, Democratic People's Republic of Korea}
\date{\today}

\begin{abstract}
The impact of local and nonlocal density-density interactions on the electronic instabilities in the honeycomb lattice is widely investigated. Some early studies proposed the emergence of interaction-induced topologically nontrivial phases, but recently, it was denied in several works including renormalization group calculations with refined momentum resolution. We use the truncated unity functional renormalization group to study the many-body instabilities of electrons on the half-filled honeycomb lattice, focusing on the effect of the exchange interaction. We show that varying the next-nearest-neighbor repulsion and nearest-neighbor exchange integral can lead to diverse ordered phases, namely, the quantum spin Hall, the spin-Kekul\'{e}, and some spin- and charge-density-wave phases. The quantum spin Hall phase can be induced by a combination of the ferromagnetic exchange and pair hopping interactions. Another exotic phase, the spin-Kekul\'{e} phase, develops in a very small region of the parameter space considered. We encounter the three-sublattice charge-density-wave phase in a large part of the parameter space. It is replaced by the incommensurate charge density wave when increasing the exchange integral. In order to reduce the computational effort, we derive the explicit symmetry relations for the bosonic propagators of the effective interaction and propose a linear-response-based approach for identifying the form factor of order parameter. Their efficiencies are confirmed by numerical calculations in our work.
\end{abstract}


\maketitle

\section{Introduction}

The experimental realization of graphene has stimulated intensive research activities related to this material. Its lattice structure, the two-dimensional honeycomb lattice, has been serving as a platform for basic research on exotic many-body phenomena \cite{ref01, ref02}. Considerable effort has been invested in numerical studies on the possible ground states of extended Hubbard models on the honeycomb lattice. Many of these works have explored a variety of possible orderings in the ground states of the systems with different interaction strengths.

Raghu {\it et al.} suggested that a topologically nontrivial quantum anomalous Hall (QAH) state could emerge on the honeycomb lattice from a large next-nearest-neighbor repulsion $V_2$ since an effective spin-orbit interaction is generated by mean-field decoupling of this term \cite{ref03}. Further mean-field studies \cite{ref04, ref05, ref06} also showed evidence of the existence of the interaction-driven QAH state of spinless fermions on the honeycomb lattice. Another type of topologically nontrivial phase emerging from only a nearest-neighbor interaction $V_1$ has been reported away from half filling \cite{ref07}.

The poorly screened Coulomb interaction in the system at half filling may also develop other conventional orders, such as bond order, charge order, and magnetic order, which then compete with the QAH. For the spinless Hubbard model which has Coulomb interactions between nearest-neighbor sites ($V_1$) and between next-nearest-neighbor sites ($V_2$), mean-field calculations proposed the existence of a Kekul\'{e} bond order phase \cite{ref04}. This phase has also been reported in Refs. \onlinecite{ref10, refB62, ref08, ref09}. Exact diagonalization (ED) \cite{ref08, ref09, ref11, ref12}, infinite density matrix renormalization group (IDMRG) \cite{ref10} studies have considered the stability of the QAH ground state and found various charge-density-wave (CDW) states competing with QAH.

For the spinful Hubbard model an on-site repulsion $U>0$ between electrons with opposite spins is added, which generates the intricate interplay between charge and spin degrees of freedom. The dominant on-site Coulomb repulsion favors an antiferromagnetic spin-density-wave (SDW) phase \cite{ref13}. The combination of on-site $U$, nearest-neighbor $V_1$, and next-nearest-neighbor $V_2$ repulsions introduces the possibility of a spinful version of QAH, i.e., the quantum spin Hall (QSH) state \cite{ref03}, and it is expected that there is a more complicated competition or coexistence of different ordering tendencies, including the conventional CDW, SDW, and QSH \cite{ref14, ref15}.

A number of recent studies of the extended Hubbard model on the honeycomb lattice by means of ED \cite{ref09, ref11}, IDMRG \cite{ref10}, functional renormalization group (FRG) \cite{ref14, ref15, ref20}, and quantum Monte Carlo \cite{ref16, ref18, ref19} indicate the suppression of the QSH or QAH by conventional ordered phases, mainly CDW, at half filling.  This discrepancy regarding the existence of the topologically nontrivial phases needs more research on possible ground states of the honeycomb lattice by more effective and credible approaches. On the other hand, many of these works have considered only the correlation effects by the Coulomb repulsion between electrons, e.g., the parameters $V_1$ and $V_2$. It is natural to expect that the inclusion of the exchange interaction between nearest-neighbor sites would give an even richer ground-state phase diagram for the honeycomb lattice at half filling.

In this work, we investigate the quantum many-body instabilities of electrons on the half-filled honeycomb lattice focusing on the effect of the nearest-neighbor exchange interaction $J$. The effect of the nearest-neighbor repulsion $V_1$ is relatively well understood. Beyond a critical coupling strength, $V_1$ destabilizes the semimetallic phase, and the system undergoes a direct and continuous quantum phase transition to a fully gapped CDW phase with opposite charge configurations on two sublattices \cite{ref20}. While the effect of parameter $V_1$ on the ground state of the half-filled honeycomb lattice was investigated in many preceding studies, the effect of $V_2$ and $J$ on it appears to be far from clear. Therefore, we use the extended Hubbard model with on-site repulsion $U$, next-nearest-neighbor repulsion $V_2$, and nearest-neighbor exchange interaction $J$ for interacting electrons on the honeycomb lattice.

The main goal of this work is to explore the existence of QSH and the effect of the exchange interaction on the electronic instabilities. To this end, we employ the recently developed truncated unity functional renormalization group (TUFRG) approach for correlated fermions \cite{ref21} with a high resolution of wave-vector dependences in the effective interaction. In addition, we present the symmetry relations for the bosonic propagators, which can reduce the computational effort in the case of the honeycomb lattice to 1/12. We also propose an efficient approach to estimate the form factor of order parameter from the TUFRG results of the effective interaction, which is based on the analysis of the linear response of the system to virtual infinitesimal external fields coupled to the fermion bilinears.

This paper is organized as follows. In Sec.~\ref{sec2}, we introduce the Hamiltonian of the extended Hubbard model for spin-1/2 fermions on the honeycomb lattice and the TUFRG approach. In Sec.~\ref{sec3}, we derive symmetry relations for the effective interactions and bosonic propagators of electrons in the system and discuss a method to estimate the form factors of various order parameters in three channels, i.e., the pairing, spin, and charge channels. In Sec.~\ref{sec4}, we present and analyze a tentative phase diagram for electrons subjected to on-site, next-nearest-neighbor repulsions and nearest-neighbor exchange interaction. Finally, in Sec.~\ref{sec5}, we draw our conclusions.

\section{Model and Method} \label{sec2}

\subsection{Extended Hubbard model}\label{sec2A}
We study spin-1/2 fermions on the honeycomb lattice at half filling which interact with each other via on-site repulsion $U$, next-nearest-neighbor repulsion $V_2$, and nearest-neighbor exchange interaction $J$. For simplicity, we neglect the nearest-neighbor repulsion $V_1$. The Hamiltonian of the extended Hubbard model is composed of a single-particle part $H_0$ and an interaction part $H_{{\rm{int}}}$,
\begin{eqnarray} \label{eq01}
H = H_0  + H_{{\rm{int}}}, 
\end{eqnarray}
where $H_0$ is described by a tight-binding Hamiltonian with nearest-neighbor hopping $t$ for the honeycomb lattice at half filling (i.e., $\mu  = 0$),
\begin{eqnarray} \label{eq02}
H_0  =  - t\sum\limits_{\left\langle {i,j} \right\rangle ,\sigma } {(c_{i,A,\sigma }^\dag  } c_{j,B,\sigma }  + {\rm{H.c.}}).
\end{eqnarray}
Here the operator $c_{i,o,\sigma }^\dag$ ($c_{i,o,\sigma }$) creates (annihilates) an electron at lattice site $i$ in the sublattice (orbital) $o$ with spin polarity $\sigma$. The unit of length is set to the lattice constant $a$, i.e., the distance between next-nearest-neighbor sites.

The interaction part of the Hamiltonian is given by
\begin{eqnarray}\label{eq03}
\begin{split}
& H_{{\rm{int}}}=U\sum\limits_{i,o} {n_{i,o, \uparrow } } n_{i,o, \downarrow }  + V_2 \sum\limits_{\scriptstyle \left\langle {\left\langle {io,jo} \right\rangle } \right\rangle  \hfill \atop 
	\scriptstyle {\kern 6pt} o,\sigma ,\sigma ' \hfill} {n_{i,o,\sigma } } n_{j,o,\sigma '}  \\ 
& {\kern 28pt} + J\sum\limits_{\scriptstyle \left\langle {iA,jB} \right\rangle  \hfill \atop 
	\scriptstyle {\kern 1pt} {\kern 1pt} {\kern 1pt} {\kern 1pt} {\kern 1pt} {\kern 1pt} \sigma ,\sigma ' \hfill} {c_{i,A,\sigma }^\dag  } c_{j,B,\sigma '}^\dag  c_{i,A,\sigma '} c_{j,B,\sigma }\\  
& {\kern 28pt} + J\sum\limits_{\left\langle {iA,jB} \right\rangle } {(c_{i,A, \uparrow }^\dag  } c_{i,A, \downarrow }^\dag  c_{j,B, \downarrow } c_{j,B, \uparrow }  + {\rm{H.c.}}),
\end{split}
\end{eqnarray}
where $n_{i,o,\sigma }=c_{i,o,\sigma }^\dag c_{i,o,\sigma }$ is the local electron density operator and the sums $\sum {_{\left\langle {iA,jB} \right\rangle } }$ and $\sum {_{\left\langle {\left\langle {io,jo} \right\rangle } \right\rangle } }$ run over nearest and next-nearest neighbors, respectively. The terms in Eq. (\ref{eq03}) describe the on-site Coulomb interaction, next-nearest-neighbor Coulomb interaction, the nearest-neighbor ferromagnetic exchange, and nearest-neighbor pair hopping. [For the real $p_z$ orbitals, the ferromagnetic exchange is accompanied by the pair hopping as in Eq. (\ref{eq03}).] Although the magnitude of $J$ is known to be rather small in realistic systems, we artificially enhance it to investigate its effect on the ground state using TUFRG.

\subsection{Truncated unity functional\\ renormalization group method}\label{sec2B}
The FRG method \cite{refA1, ref22} was developed as an unbiased tool for detecting many-body instabilities in interacting Fermi systems. This method has been successfully applied to a number of unconventional superconductors such as cuprates \cite{ref23, refB46, refB47, refB48, refB49}, iron pnictides \cite{ref24, refB52, ref33, refB55, refB56}, and strontium ruthenate \cite{refA2}. It has also been employed to identify the leading ordering tendencies in the honeycomb lattice \cite{refB58, ref26, ref34, refB60, refB62, ref14, ref15, ref20} and bilayer \cite{refB64, refB65, refB66} and trilayer \cite{refB67} honeycomb lattices. Comprehensive introductions to the FRG method can be found in Refs. \onlinecite{refA5, refA3, refA4}.

The TUFRG method \cite{ref21, ref14} is a modification of the FRG method. It is based on prior channel-decomposed FRG \cite{ref25} and singular-mode FRG \cite{ref26} schemes. The advantage of the TUFRG is the increased momentum resolution in the low-energy effective interaction achieved by the numerically efficient scheme. We use the orbital picture of TUFRG, which ensures excellent convergence in the expansion of the effective interaction. The $\Omega$ scheme \cite{ref25} is employed as the regulator for infrared divergences. In this scheme the bare propagator $G_{o_1 o_2 }^0 (\omega ,{\bf{k}})$ for Matsubara frequency $\omega$, wave vector $\bf{k}$, and orbital indices $o_1 ,o_2$ (for the honeycomb lattice, the orbital index means sublattice $A$ or $B$) gets modified with energy scale $\Omega$, i.e.,
\begin{equation}
\nonumber
G_{o_1 o_2 }^0 (\omega ,{\bf{k}}) \to G_{o_1 o_2 }^{0,\Omega } (\omega ,{\bf{k}}) = \frac{{\omega ^2 }}{{\omega ^2  + \Omega ^2 }}G_{o_1 o_2 }^0 (\omega ,{\bf{k}}).
\end{equation}
The modified propagator $G^{0,\Omega }$ is then used to set up the expression for the generating functional of one-particle-irreducible vertex functions, which is now scale dependent as well, $\Gamma  \to \Gamma ^\Omega$. The functional flow equation is generated by differentiating $\Gamma ^\Omega$ with respect to $\Omega$, which produces a hierarchy of flow equations for the vertex functions. For our analysis we use a truncation in which all $n$-particle vertices with $n \ge 3$ and self-energy feedback are neglected. Such an approximation has been proved to be suitable in weak-coupling regimes \cite{ref22}. For spin-SU(2)-invariant systems, the two-particle part of the generating functional is expressed in terms of the two-particle vertex functions (i.e., the effective interactions) $V^\Omega$ and the Grassmann variables $\bar\psi , \psi$ as follows:
\begin{widetext}
\begin{eqnarray}
\nonumber\Gamma ^{\Omega ,(4)} [\bar \psi ,\psi ] = \frac{1}{2}\int {d\xi _1 }  \cdots d\xi _4 {\kern 1pt} {\kern 1pt} V_{o_1 o_2 ,o_3 o_4 }^\Omega  (k_1 ,k_2 ;k_3 ,k_4 )
\delta (k_1  + k_2  - k_3  - k_4 )\sum\limits_{\sigma ,\sigma '} {\bar \psi _\sigma  } (\xi _1 )\bar \psi _{\sigma '} (\xi _2 )\psi _{\sigma '} (\xi _4 )\psi _\sigma  (\xi _3 ),
\end{eqnarray}
where $k_i  = (\omega _i ,{\bf{k}}_i )$ and $\xi _i  = (\omega _i ,{\bf{k}}_i ,o_i )$ are multi-indices gathering a Matsubara frequency $\omega _i$, wave vector ${\bf{k}}_i$, and orbital index $o_i$ and $\int {d\xi _i }$ stands for the notation $\int {\frac{{d{\bf{k}}_i }}{{S_{BZ} }}\frac{1}{\beta }} \sum {_{\omega _i } } \sum {_{o_i } }$ with the Brillouin zone area $S_{BZ}$ and inverse temperature $\beta$. Since the most singular part of the effective interaction comes from the zero frequency and we are interested in ground-state properties, we will neglect its frequency dependence, with external frequencies set to zero. This approximation has proven to give reliable results in a lot of two-dimensional correlated systems \cite{refA3, refA4}. In the approximation neglecting frequency dependence, the flow equation for the effective interaction consists of three contributions \cite{ref22},
\begin{eqnarray}\label{eq05}
\begin{split}
&\frac{d}{{d\Omega }}V_{o'_1 o'_2 ,o_1 o_2 }^\Omega ({\bf{k}}'_1 ,{\bf{k}}'_2 ; {\bf{k}}_1 ,{\bf{k}}_2 ) =
 J_{o'_1 o'_2 ,o_1 o_2 }^{{\rm{pp}}} ({\bf{k}}'_1 ,{\bf{k}}'_2 ;{\bf{k}}_1 ,{\bf{k}}_2 )+
 J_{o'_1 o'_2 ,o_1 o_2 }^{{\rm{ph,}} {\rm{cr}}} ({\bf{k}}'_1 ,{\bf{k}}'_2 ;{\bf{k}}_1 ,{\bf{k}}_2 ) +
 J_{o'_1 o'_2 ,o_1 o_2 }^{{\rm{ph,}} {\rm{d}}} ({\bf{k}}'_1 ,{\bf{k}}'_2 ;{\bf{k}}_1 ,{\bf{k}}_2 ),
\end{split}
\end{eqnarray}
where the particle-particle, crossed, and direct particle-hole contributions read
\begin{eqnarray}\label{eq06}
\begin{split}
& J_{o'_1 o'_2 ,o_1 o_2 }^{{\rm{pp}}} ({\bf{k}}'_1 ,{\bf{k}}'_2 ;{\bf{k}}_1 ,{\bf{k}}_2 ) =  - \sum\limits_{\mu ,\mu '} {\sum\limits_{\nu ,\nu '} {\int {dp} } } {\kern 1pt} {\kern 1pt} \frac{d}{{d\Omega }}[G_{\mu \nu }^{0,\Omega } (\omega ,{\bf{p}} + {\bf{k}}'_1  + {\bf{k}}'_2 )G_{\mu '\nu '}^{0,\Omega } ( - \omega , - {\bf{p}})]\\
&{\kern 50pt}\times V_{o'_1 o'_2 ,\mu \mu '}^\Omega  ({\bf{k}}'_1 ,{\bf{k}}'_2 ;{\bf{p}} + {\bf{k}}'_1  + {\bf{k}}'_2 , - {\bf{p}}) \cdot V_{\nu \nu ',o_1 o_2 }^\Omega  ({\bf{p}} + {\bf{k}}'_1  + {\bf{k}}'_2 , - {\bf{p}};{\bf{k}}_1 ,{\bf{k}}_2 ),\\
&J_{o'_1 o'_2 ,o_1 o_2 }^{{\rm{ph,cr}}} ({\bf{k}}'_1 ,{\bf{k}}'_2 ;{\bf{k}}_1 ,{\bf{k}}_2 ) =  - \sum\limits_{\mu ,\mu '} {\sum\limits_{\nu ,\nu '} {\int {dp} } } {\kern 1pt} {\kern 1pt} \frac{d}{{d\Omega }}[G_{\mu \nu }^{0,\Omega } (\omega ,{\bf{p}} + {\bf{k}}'_1  - {\bf{k}}_2 )G_{\nu '\mu '}^{0,\Omega } (\omega ,{\bf{p}})]\\
&{\kern 50pt}\times V_{o'_1 \mu ',\mu o_2 }^\Omega  ({\bf{k}}'_1 ,{\bf{p}};{\bf{p}} + {\bf{k}}'_1  - {\bf{k}}_2 ,{\bf{k}}_2 ) \cdot V_{\nu o'_2 ,o_1 \nu '}^\Omega  ({\bf{p}} + {\bf{k}}'_1  + {\bf{k}}'_2 , - {\bf{p}};{\bf{k}}_1 ,{\bf{k}}_2 ),\\
&J_{o'_1 o'_2 ,o_1 o_2 }^{{\rm{ph,d}}} ({\bf{k}}'_1 ,{\bf{k}}'_2 ;{\bf{k}}_1 ,{\bf{k}}_2 ) =  - \sum\limits_{\mu ,\mu '} {\sum\limits_{\nu ,\nu '} {\int {dp} } } {\kern 1pt} {\kern 1pt} \frac{d}{{d\Omega }}[G_{\mu \nu }^{0,\Omega } (\omega ,{\bf{p}} + {\bf{k}}'_1  - {\bf{k}}_1 )G_{\nu '\mu '}^{0,\Omega } (\omega ,{\bf{p}})]\\
&{\kern 50pt}\times [V_{o'_1 \mu ',\mu o_1 }^\Omega  ({\bf{k}}'_1 ,{\bf{p}};{\bf{p}} + {\bf{k}}'_1  - {\bf{k}}_1 ,{\bf{k}}_1 ) \cdot V_{\nu o'_2 ,\nu 'o_2 }^\Omega  ({\bf{p}} + {\bf{k}}'_1  - {\bf{k}}_1 ,{\bf{k}}'_2 ;{\bf{p}},{\bf{k}}_2 )\\
&{\kern 50pt} + V_{o'_1 \mu ',o_1 \mu }^\Omega  ({\bf{k}}'_1 ,{\bf{p}};{\bf{k}}_1 ,{\bf{p}} + {\bf{k}}'_1  - {\bf{k}}_1 ) \cdot V_{\nu o'_2 ,o_2 \nu '}^\Omega  ({\bf{p}} + {\bf{k}}'_1  - {\bf{k}}_1 ,{\bf{k}}'_2 ;{\bf{k}}_2 ,{\bf{p}})\\
&{\kern 50pt} - 2V_{o'_1 \mu ',o_1 \mu }^\Omega  ({\bf{k}}'_1 ,{\bf{p}};{\bf{k}}_1 ,{\bf{p}} + {\bf{k}}'_1  - {\bf{k}}_1 ) \cdot V_{\nu o'_2 ,\nu 'o_2 }^\Omega  ({\bf{p}} + {\bf{k}}'_1  - {\bf{k}}_1 ,{\bf{k}}'_2 ;{\bf{p}},{\bf{k}}_2 )],
\end{split}
\end{eqnarray}
with the shorthand notation $\int {dp}  = \int {\frac{{d{\bf{p}}}}{{S_{BZ} }}\frac{1}{\beta }} \sum {_\omega  }$ and implicit constraint ${\bf{k}}'_1  + {\bf{k}}'_2  = {\bf{k}}_1  + {\bf{k}}_2$.
Below we explain the main algorithm of the TUFRG, pointing out a small difference from Ref. \onlinecite{ref14}.

The effective interactions can be expressed by a decomposition into single-channel coupling functions
\begin{eqnarray}\label{eq07}
\begin{split}
V_{o'_1 o'_2 ,o_1 o_2 }^\Omega  ({\bf{k}}'_1 ,{\bf{k}}'_2 ;{\bf{k}}_1 ,{\bf{k}}_2 ) =  V_{o'_1 o'_2 ,o_1 o_2 }^{(0)} ({\bf{k}}'_1 ,{\bf{k}}'_2 ;{\bf{k}}_1 ,{\bf{k}}_2 ) + \Phi _{o'_1 o'_2 ,o_1 o_2 }^{{\rm{SC}}} ({\bf{k}}'_1  + {\bf{k}}'_2 ; - {\bf{k}}'_2 , - {\bf{k}}_2 )\\
 + \Phi _{o'_1 o'_2 ,o_1 o_2 }^{\rm{C}} ({\bf{k}}'_1  - {\bf{k}}_2 ;{\bf{k}}_2 ,{\bf{k}}'_2 ) + \Phi _{o'_1 o'_2 ,o_1 o_2 }^{\rm{D}} ({\bf{k}}'_1  - {\bf{k}}_1 ;{\bf{k}}_1 ,{\bf{k}}'_2 ),
\end{split}
\end{eqnarray}
where $V^{(0)}$ is the initial bare interaction and is obtained by the Fourier transform of the interaction Hamiltonian in Eq. (\ref{eq03}). Equation (\ref{eq07}) has no momentum like $({\bf{k}}_1  - {\bf{k}}_2 )/2$ in Eq. (5) of Ref. \onlinecite{ref14}, which ensures the periodicity of the reciprocal lattice for the $\Phi$ functions. The coupling functions $\Phi$ are developed during the flow according to
\begin{eqnarray}\label{eq08}
\begin{split}
&\frac{d}{{d\Omega }}\Phi _{o'_1 o'_2 ,o_1 o_2 }^{{\rm{SC}}} ({\bf{k}}'_1  + {\bf{k}}'_2 ; - {\bf{k}}'_2 , - {\bf{k}}_2 ) = J_{o'_1 o'_2 ,o_1 o_2 }^{{\rm{pp}}} ({\bf{k}}'_1 ,{\bf{k}}'_2 ;{\bf{k}}_1 ,{\bf{k}}_2 ),\\
&\frac{d}{{d\Omega }}\Phi _{o'_1 o'_2 ,o_1 o_2 }^{\rm{C}} ({\bf{k}}'_1  - {\bf{k}}_2 ;{\bf{k}}_2 ,{\bf{k}}'_2 ) = J_{o'_1 o'_2 ,o_1 o_2 }^{{\rm{ph,}}{\kern 1pt} {\kern 1pt} {\rm{cr}}} ({\bf{k}}'_1 ,{\bf{k}}'_2 ;{\bf{k}}_1 ,{\bf{k}}_2 ),\\
&\frac{d}{{d\Omega }}\Phi _{o'_1 o'_2 ,o_1 o_2 }^{\rm{D}} ({\bf{k}}'_1  - {\bf{k}}_1 ;{\bf{k}}_1 ,{\bf{k}}'_2 ) = J_{o'_1 o'_2 ,o_1 o_2 }^{{\rm{ph,}}{\kern 1pt} {\kern 1pt} {\rm{d}}} ({\bf{k}}'_1 ,{\bf{k}}'_2 ;{\bf{k}}_1 ,{\bf{k}}_2 ).\\
\end{split}
\end{eqnarray}
Taking into account the strong dependence on transfer momenta (first argument) and weak dependence on other momenta, one can expand the coupling functions $\Phi$ in a suitable scale-independent basis $f_m$ and three bosonic propagators $P, C, D$,
\begin{eqnarray}\label{eq09}
\begin{split}
&{\kern 130pt} \Phi _{o'_1 o'_2 ,o_1 o_2 }^{{\rm{SC}}} ({\bf{q}}; {\bf{p}},{\bf{k}}) = 
\sum\limits_{m,n} {P_{o'_1 o'_2 m,o_1 o_2 n} } ({\bf{q}})f_m^* ({\bf{p}})f_n ({\bf{k}}),\\
&\Phi _{o'_1 o'_2 ,o_1 o_2 }^{\rm{C}} ({\bf{q}}; {\bf{p}},{\bf{k}}) =
\sum\limits_{m,n} {C_{o'_1 o_2 m,o_1 o'_2 n} } ({\bf{q}})f_m^* ({\bf{p}})f_n ({\bf{k}}),
\Phi _{o'_1 o'_2 ,o_1 o_2 }^{\rm{D}} ({\bf{q}};  {\bf{p}},{\bf{k}}) =
\sum\limits_{m,n} {D_{o'_1 o_1 m,o_2 o'_2 n} } ({\bf{q}})f_m^* ({\bf{p}})f_n ({\bf{k}}).
\end{split}
\end{eqnarray}
Note that the $P, C, D$ matrices have different sequences of the orbital indices in Eq. (\ref{eq09}). In practical computation the infinite basis has to be truncated, and the truncation error is generally smaller for the orbital picture than the one for the band picture. Then the coupling functions are represented by the three matrices $P$, $C$, and $D$, each having just one momentum dependence. This enables us to perform a calculation with high momentum resolution. The combination of Eqs. (\ref{eq06})-(\ref{eq09}) yields the flow equation for the bosonic propagators containing intricate terms in which internal bosonic propagators appear in the fermionic loops and have to be integrated out, which is challenging in calculations. With insertions of truncated partitions of unity the fermionic propagators are decoupled from the bosonic propagators, yielding a very simplified flow equation. The ultimate flow equation for the bosonic propagators reads
\begin{eqnarray}\label{eq10}
\begin{split}
&\frac{{dP({\bf{q}})}}{{d\Omega }} = V^{\rm P} ({\bf{q}})\dot \chi ^{{\rm{pp}}} ({\bf{q}})V^{\rm P} ({\bf{q}}),{\kern 1pt} {\kern 1pt} {\kern 1pt} {\kern 1pt} {\kern 1pt} \frac{{dC({\bf{q}})}}{{d\Omega }} = V^{\rm C} ({\bf{q}})\dot \chi ^{{\rm{ph}}} ({\bf{q}})V^{\rm C} ({\bf{q}}),\\
&\frac{{dD({\bf{q}})}}{{d\Omega }} = [V^{\rm C} ({\bf{q}}) - V^{\rm D} ({\bf{q}})]\dot \chi ^{{\rm{ph}}} ({\bf{q}})V^{\rm D} ({\bf{q}}) + V^{\rm D} ({\bf{q}})\dot \chi ^{{\rm{ph}}} ({\bf{q}})[V^{\rm C} ({\bf{q}}) - V^{\rm D} ({\bf{q}})],
\end{split}
\end{eqnarray}
where
\begin{eqnarray}\label{eq11}
\begin{split}
&\dot \chi _{o'_1 o'_2 m,o_1 o_2 n}^{{\rm{pp}}} ({\bf{q}}) =  - \int {dk} \frac{d}{{d\Omega }}[G_{o'_1 o_1 }^{0,\Omega } (\omega ,{\bf{k}} + {\bf{q}})G_{o'_2 o_2 }^{0,\Omega } ( - \omega , - {\bf{k}})]f_m ({\bf{k}})f_n^* ({\bf{k}}),\\
&\dot \chi _{o'_1 o'_2 m,o_1 o_2 n}^{{\rm{ph}}} ({\bf{q}}) =  - \int {dk} \frac{d}{{d\Omega }}[G_{o'_1 o_1 }^{0,\Omega } (\omega ,{\bf{k}} + {\bf{q}})G_{o_2 o'_2 }^{0,\Omega } (\omega ,{\bf{k}})]f_m ({\bf{k}})f_n^* ({\bf{k}}),
\end{split}
\end{eqnarray}
and $V^{\rm {P,C,D}}$ are projections of the effective interactions onto the form of the three channels,
\begin{eqnarray}\label{eq12}
\begin{split}
&V_{o'_1 o'_2 m,o_1 o_2 n}^{\rm P} ({\bf{q}}) = \frac{1}{{S_{BZ}^2 }}\int {d{\bf{p}}} \int {d{\bf{p'}}} f_m ({\bf{p}})f_n^* ({\bf{p'}})V_{o'_1 o'_2 ,o_1 o_2 }^\Omega  ({\bf{p}} + {\bf{q}}, - {\bf{p}};{\bf{p'}} + {\bf{q}}, - {\bf{p'}}),\\
&V_{o'_1 o_2 m,o_1 o'_2 n}^{\rm C} ({\bf{q}}) = \frac{1}{{S_{BZ}^2 }}\int {d{\bf{p}}} \int {d{\bf{p'}}} f_m ({\bf{p}})f_n^* ({\bf{p'}})V_{o'_1 o'_2 ,o_1 o_2 }^\Omega  ({\bf{p}} + {\bf{q}},{\bf{p'}};{\bf{p'}} + {\bf{q}},{\bf{p}}),\\
&V_{o'_1 o_1 m,o_2 o'_2 n}^{\rm D} ({\bf{q}}) = \frac{1}{{S_{BZ}^2 }}\int {d{\bf{p}}} \int {d{\bf{p'}}} f_m ({\bf{p}})f_n^* ({\bf{p'}})V_{o'_1 o'_2 ,o_1 o_2 }^\Omega  ({\bf{p}} + {\bf{q}},{\bf{p'}};{\bf{p}},{\bf{p'}} + {\bf{q}}).
\end{split}
\end{eqnarray}
The bare propagator $G_{oo'}^0 (\omega ,{\bf{k}})$ in the orbital picture is
\begin{eqnarray}\label{eq13}
G_{oo'}^0 (\omega ,{\bf{k}}) = \sum\limits_b {T_{ob} ({\bf{k}})} T_{o'b}^* ({\bf{k}})
\left[ {i\omega  - E_b ({\bf{k}})}\right]^{ - 1}.
\end{eqnarray}
Here $E_b ({\bf{k}})$ is the one-particle energy of the electron with the wave vector $\bf{k}$ in band $b$, and $T_{ob} ({\bf{k}})$ is the element of the transformation matrix between the annihilation operators in the orbital picture $c_{{\bf{k}},o,\sigma }$ and ones in the band picture $b_{{\bf{k}},b,\sigma}$, namely, $c_{{\bf{k}},o,\sigma }  = \sum{_b} {\kern 2pt} T_{ob}({\bf{k}}) b_{{\bf{k}},b,\sigma }$.
By substituting Eqs. (\ref{eq07}) and (\ref{eq09}) into Eq. (\ref{eq12}), one can represent the projections of the effective interactions of Eq. (\ref{eq12}) in terms of three matrices, $P, C$, and $D$, thus obtaining the expression of the flow equation (\ref{eq10}) via only these matrices. When we choose the plane wave, $f_m ({\bf{p}}) = e^{i{\bf{R}}_m  \cdot {\bf{p}}}$, as the basis, Eq. (\ref{eq12}) gives
\begin{eqnarray}\label{eq15}
\begin{split}
& V^{\rm P} ({\bf{q}}) = V^{\rm P,(0)} ({\bf{q}}) + P({\bf{q}}) + V^{\rm P,(C)} ({\bf{q}}) + V^{\rm P,(D)} ({\bf{q}}),\\
& V_{o'_1 o'_2 m,o_1 o_2 n}^{\rm P,(C)} ({\bf{q}}) = \sum\limits_l {\tilde C_{o'_1 ,o_2 ,{\bf{R}}_l ;o_1 ,o'_2 ,{\bf{R}}_m  + {\bf{R}}_n  - {\bf{R}}_l } } ({\bf{R}}_n  - {\bf{R}}_l )e^{i({\bf{R}}_n  - {\bf{R}}_l ) \cdot {\bf{q}}},\\ 
& V_{o'_1 o'_2 m,o_1 o_2 n}^{\rm P,(D)} ({\bf{q}}) = \sum\limits_l {\tilde D_{o'_1 ,o_1 ,{\bf{R}}_l ;o_2 ,o'_2 ,{\bf{R}}_m  - {\bf{R}}_n  - {\bf{R}}_l } } ( - {\bf{R}}_n  - {\bf{R}}_l )e^{ - i{\bf{R}}_l  \cdot {\bf{q}}}, 
\end{split}
\end{eqnarray}
\begin{eqnarray}\label{eq16}
\begin{split}
&V^{\rm C} ({\bf{q}}) = V^{\rm C,(0)} ({\bf{q}}) + C({\bf{q}}) + V^{\rm C,(P)} ({\bf{q}}) + V^{\rm C,(D)} ({\bf{q}}),\\
&V_{o'_1 o_2 m,o_1 o'_2 n}^{\rm C,(P)} ({\bf{q}}) = \sum\limits_l {\tilde P_{o'_1 ,o'_2 ,{\bf{R}}_l ;o_1 ,o_2 ,{\bf{R}}_m  + {\bf{R}}_n  - {\bf{R}}_l } ({\bf{R}}_n  - {\bf{R}}_l )e^{i({\bf{R}}_n  - {\bf{R}}_l ) \cdot {\bf{q}}} },\\
&V_{o'_1 o_2 m,o_1 o'_2 n}^{\rm C,(D)} ({\bf{q}}) = \sum\limits_l {\tilde D_{o'_1 ,o_1 ,{\bf{R}}_l ;o_2 ,o'_2 ,{\bf{R}}_n  + {\bf{R}}_l  - {\bf{R}}_m } ( - {\bf{R}}_m )e^{ - i{\bf{R}}_l  \cdot {\bf{q}}} },
\end{split}
\end{eqnarray}
\begin{eqnarray}\label{eq17}
\begin{split}
&V^{\rm D} ({\bf{q}}) = V^{\rm D,(0)} ({\bf{q}}) + D({\bf{q}}) + V^{\rm D,(P)} ({\bf{q}}) + V^{\rm D,(C)} ({\bf{q}}),\\
&V_{o'_1 o_1 m,o_2 o'_2 n}^{\rm D,(P)} ({\bf{q}}) = \sum\limits_l {\tilde P_{o'_1 ,o'_2 ,{\bf{R}}_l ;o_1 ,o_2 ,{\bf{R}}_l  - {\bf{R}}_m  - {\bf{R}}_n } ( - {\bf{R}}_m )e^{i({\bf{R}}_n  - {\bf{R}}_l ) \cdot {\bf{q}}} },\\
&V_{o'_1 o_1 m,o_2 o'_2 n}^{\rm D,(C)} ({\bf{q}}) = \sum\limits_l {\tilde C_{o'_1 ,o_2 ,{\bf{R}}_l ;o_1 ,o'_2 ,{\bf{R}}_n  + {\bf{R}}_l  - {\bf{R}}_m } ( - {\bf{R}}_m )e^{ - i{\bf{R}}_l  \cdot {\bf{q}}} },
\end{split}
\end{eqnarray}
where $\tilde P({\bf{R}}_m)$, $\tilde C({\bf{R}}_m)$, and $\tilde D({\bf{R}}_m)$ are the Fourier transforms of matrices $P({\bf{q}}),C({\bf{q}}),D({\bf{q}})$. In Eqs. (\ref{eq15})-(\ref{eq17}), $V^{\rm P,(0)}({\bf{q}})$, $V^{\rm C,(0)}({\bf{q}})$, and $V^{\rm D,(0)}({\bf{q}})$, which are obtained by replacing $V^\Omega$ with $V^{(0)}$ in Eq. (\ref{eq12}), are the projections of the initial effective interactions $V_{o'_1 o'_2 ,o_1 o_2 }^{(0)} ({\bf{k}}'_1 ,{\bf{k}}'_2 ;{\bf{k}}_1 ,{\bf{k}}_2 )$ [i.e., the Fourier transform of the interaction Hamiltonian in Eq. (\ref{eq03})] onto the three channels.
\end{widetext}

\section{Symmetries and Order Parameters}\label{sec3}

\subsection{Symmetries of bosonic propagators}\label{sec3A}
\begin{figure}[h!]
	\begin{center}
		\includegraphics[width=8.6cm]{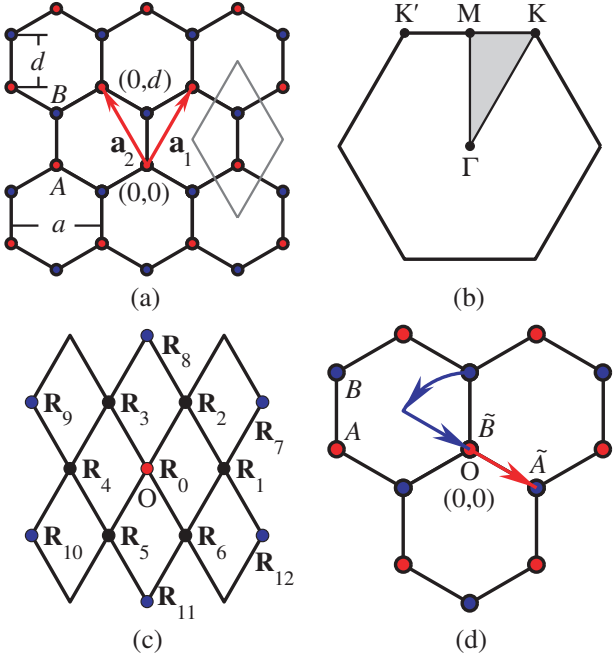}
	\end{center}
	\caption{(Color online) Honeycomb lattice, its Brillouin zone, Bravais lattice vectors in the bases, and illustration of $\pi/3$ rotation. (a) Lattice structure. The $A$ ($B$) sublattice is indicated by red (blue) spheres. The primitive vectors of the lattice are ${\bf{a}}_1$ and ${\bf{a}}_2$, and a unit cell is shown by the part enclosed with a gray line. (b) Brillouin zone. The gray part is the irreducible region of the Brillouin zone. (c) Bravais lattice vectors ${\bf{R}}_m$ in the 13 plane-wave bases $f_m ({\bf{p}}) = e^{i{\bf{R}}_m  \cdot {\bf{p}}}$ used by us. (d) Transfer of $A, B$ sites upon $\pi/3$ rotation followed by appropriate shift.}
	\label{fig1}
\end{figure}

The structure of the honeycomb lattice is shown in Fig. \ref{fig1}(a). The lattice has $C_{6v}$ symmetry with respect to the center of hexagons. This symmetry yields the symmetry relations between the bosonic propagators with different momentum transfers. By these relations, the bosonic propagators with transfer momenta in the whole Brillouin zone (BZ) can be generated from ones with transfer momenta in the irreducible region of the BZ [see Fig. \ref{fig1}(b)], thus reducing the computational effort to $1/12$. The symmetry relations are rather complicated in the orbital picture, but when using the plane-wave basis, we can find the explicit symmetry relations for the bosonic propagators. For the expansion of the coupling functions in Eq. (\ref{eq09}), we use 13 plane-wave bases with Bravais lattice vectors ${\bf{R}}_m$ shown in Fig. \ref{fig1}(c), which are sufficient in the orbital picture.

The Bloch sum of orbitals $o$ is expressed as
\begin{eqnarray}\label{eq18}
\begin{split}
\Psi _{{\bf{k}},o} ({\bf{r}})& = \sum\limits_i {\Psi _{i,o} } ({\bf{r}})e^{i{\bf{k}} \cdot {\bf{R}}_i } /\sqrt N\\
& = \sum\limits_i {\Phi _\pi  } ({\bf{r}} - {\bf{R}}_i  - {\bf{d}}_o )e^{i{\bf{k}} \cdot {\bf{R}}_i } /\sqrt N,
\end{split}
\end{eqnarray}
where $\Phi_\pi,{\bf{R}}_i ,{\bf{d}}_o$ are the $\pi$-orbital wave function, the Bravais lattice vector of the unit cell $i$, and the relative position of the sublattice $o$, respectively. Under a symmetry operation $\hat G = (Q|{\bf{t}})$, which is a rotation $Q$ followed by shift $\bf{t}$, the function $\Psi _{{\bf{k}},o} ({\bf{r}})$ is transformed to
\begin{equation}\label{eq19}
\hat G\Psi _{{\bf{k}},o} ({\bf{r}}) = \exp ( - iQ{\bf{k}} \cdot {\bf{u}}_o )\Psi _{Q{\bf{k}},\tilde o} ({\bf{r}}).
\end{equation}
Here the Bravais lattice vector ${\bf{u}}_o$ and the orbital index $\tilde o$ are determined by
\begin{equation}\label{eq20}
Q{\bf{d}}_o  + {\bf{t}} = {\bf{u}}_o  + {\bf{d}}_{\tilde o},
\end{equation}
which means that the atom of sublattice $o$ in the unit cell with the origin is transferred to the site of sublattice $\tilde o$ in the unit cell at the position ${\bf{u}}_o$ [e.g., for $\pi/3$ rotation, see Fig. \ref{fig1}(d)]. The values of indices $\tilde o$ and vectors ${\bf{u}}_o$ for representative operations are shown in Table \ref{tab01}.
\begin{table}[h!]
	\caption{Sublattice indices $\tilde o$ and Bravais lattice vectors ${\bf{u}}_o$ for the representative symmetry operations $\hat G = (Q|{\bf{t}})$. All other symmetry operations can be represented by a combination of these operations and translations. The vectors ${\bf{R}}_m$ are shown in Fig. \ref{fig1}(c).}
	\begin{center}
		\begin{tabular}{|c|c|c|c|c|c|}
			\hline
			$Q$ (at the origin) & $\bf{t}$ & ${\bf{u}}_A$ & ${\bf{u}}_B$ & $\tilde A$ & $\tilde B$\\
			\hline
			Reflection $x\to-x$ & $(0,0)$ & ${\bf{R}}_0$ & ${\bf{R}}_0$ & $A$ & $B$\\
			$\pi/3$ rotation & $d(\sqrt3/2,-1/2)$ & ${\bf{R}}_6$ & ${\bf{R}}_0$ & $B$ & $A$\\
			$-\pi/3$ rotation & $-d(\sqrt3/2,1/2)$ & ${\bf{R}}_5$ & ${\bf{R}}_0$ & $B$ & $A$\\
			$\pi$ rotation & $d(1,0)$ & ${\bf{R}}_0$ & ${\bf{R}}_0$ & $B$ & $A$\\
			\hline
		\end{tabular}
	\end{center}
	\label{tab01}
\end{table}

Equation (\ref{eq19}) is equivalent to the transformation
\begin{eqnarray}
\hat G:c_{{\bf{k}},\tilde o,\sigma }  \to \tilde c_{{\bf{k}},\tilde o,\sigma }  = e^{ - i{\bf{k}} \cdot {\bf{u}}_o } c_{Q^{ - 1} {\bf{k}},o,\sigma }.
\end{eqnarray}
This equation gives
\begin{eqnarray}
\begin{split}
\hat G:& \sum\limits_{\tilde o_1 ,\tilde o_2 ,\tilde o_3 ,\tilde o_4 } {\int {dk_1 } }  \cdots dk_4  V_{\tilde o_1 \tilde o_2 ,\tilde o_3 \tilde o_4 }^\Omega  (k_1 ,k_2 ;k_3 ,k_4 )\\
& {\kern 20pt} \times \bar \psi _\sigma  (k_1 ,\tilde o_1 )\bar \psi _{\sigma '} (k_2 ,\tilde o_2 )\psi _{\sigma '} (k_4 ,\tilde o_4 )\psi _\sigma  (k_3 ,\tilde o_3 )\\
\to & \sum\limits_{o_1 ,o_2 ,o_3 ,o_4 } {\int {dk_1 } }  \cdots dk_4 {\kern 1pt} {\kern 1pt} V_{\tilde o_1 \tilde o_2 ,\tilde o_3 \tilde o_4 }^\Omega  (Qk_1 ,Qk_2 ;Qk_3 ,Qk_4 )\\
& {\kern 20pt} \times e^{iQ{\bf{k}}_1  \cdot {\bf{u}}_{o_1 } } e^{iQ{\bf{k}}_2  \cdot {\bf{u}}_{o_2 } } e^{ - iQ{\bf{k}}_4  \cdot {\bf{u}}_{o_4 } } e^{ - iQ{\bf{k}}_3  \cdot {\bf{u}}_{o_3 } }\\ 
& {\kern 20pt}\times \bar \psi _\sigma  (k_1 ,o_1 )\bar \psi _{\sigma '} (k_2 ,o_2 )\psi _{\sigma '} (k_4 ,o_4 )\psi _\sigma  (k_3 ,o_3 ),
\nonumber
\end{split}
\end{eqnarray}
which is equivalent to the following transformation:
\begin{eqnarray}\label{eq23}
\begin{split}
\hat G: & V_{o_1 o_2 ,o_3 o_4 }^\Omega  (k_1 ,k_2 ;k_3 ,k_4 ) \\ 
& \to V_{\tilde o_1 \tilde o_2 ,\tilde o_3 \tilde o_4 }^\Omega  (Qk_1 ,Qk_2 ;Qk_3 ,Qk_4 )\\
& {\kern 15pt} \times e^{iQ{\bf{k}}_1  \cdot {\bf{u}}_{o_1 } } e^{iQ{\bf{k}}_2  \cdot {\bf{u}}_{o_2 } } e^{ - iQ{\bf{k}}_3  \cdot {\bf{u}}_{o_3 } } e^{ - iQ{\bf{k}}_4  \cdot {\bf{u}}_{o_4 } }. 
\nonumber
\end{split}
\end{eqnarray}
Thus, from the invariance of the generating functional under the point-group operation $\hat G$, the following symmetry relations for the effective interactions are obtained in the approximation neglecting the frequency dependence:
\begin{eqnarray}\label{eq24}
\begin{split}
& V_{\tilde o_1 \tilde o_2 ,\tilde o_3 \tilde o_4 }^\Omega (Q{\bf{k}}_1 ,Q{\bf{k}}_2 ;Q{\bf{k}}_3 ,Q{\bf{k}}_4 ) \\
& {\kern 15pt} = V_{o_1 o_2 ,o_3 o_4 }^\Omega  ({\bf{k}}_1 ,{\bf{k}}_2 ;{\bf{k}}_3 ,{\bf{k}}_4 ) \\
& {\kern 27pt} \times e^{ - iQ{\bf{k}}_1  \cdot {\bf{u}}_{o_1 } } e^{ - iQ{\bf{k}}_2  \cdot {\bf{u}}_{o_2 } } e^{iQ{\bf{k}}_3  \cdot {\bf{u}}_{o_3 } } e^{iQ{\bf{k}}_4  \cdot {\bf{u}}_{o_4 } }.
\end{split}
\end{eqnarray}
Combining these relations with Eq. (\ref{eq12}) and using the properties of the plane-wave bases, one can derive the symmetry relations for the projections $V^{\rm P,C,D}$. The symmetry relations for the bosonic propagators $P, C, D$ have the same form as for the projections $V^{\rm P,C,D}$. After tedious calculation, we get the following symmetry relations:
\begin{eqnarray}\label{eq25}
\nonumber &P({\rm{or}} \; C,D)_{\tilde o_1 ,\tilde o_2 ,Q{\bf{R}}_m  + {\bf{u}}_{o_1 }  - {\bf{u}}_{o_2 } ;\tilde o_3 ,\tilde o_4 ,Q{\bf{R}}_n  + {\bf{u}}_{o_3 }  - {\bf{u}}_{o_4 } } (Q{\bf{q}})\\
& = e^{ - iQ{\bf{q}} \cdot ({\bf{u}}_{o_1 }  - {\bf{u}}_{o_3 } )}  \cdot P({\rm{or}} \; C,D)_{o_1 o_2 m,o_3 o_4 n} ({\bf{q}}).
\end{eqnarray}

On the other hand, the effective interactions have particle-hole symmetry (PHS) and the remnant of antisymmetry (RAS) of Grassmann variables \cite{ref25}:
\begin{eqnarray}\label{eq26}
\begin{split}
&V_{o_1 o_2 ,o_3 o_4 }^\Omega  ({\bf{k}}_1 ,{\bf{k}}_2 ;{\bf{k}}_3 ,{\bf{k}}_4 ) \\
&{\kern 20pt} =[V_{o_3 o_4 ,o_1 o_2 }^\Omega  ({\bf{k}}_3 ,{\bf{k}}_4 ;{\bf{k}}_1 ,{\bf{k}}_2 )]^*
\end{split}
\end{eqnarray}
\begin{eqnarray}\label{eq27}
\begin{split}
&V_{o_1 o_2 ,o_3 o_4 }^\Omega  ({\bf{k}}_1 ,{\bf{k}}_2 ;{\bf{k}}_3 ,{\bf{k}}_4 ) \\
&{\kern 20pt} =V_{o_2 o_1 ,o_4 o_3 }^\Omega  ({\bf{k}}_2 ,{\bf{k}}_1 ;{\bf{k}}_4 ,{\bf{k}}_3 )
\end{split}
\end{eqnarray}
for PHS and RAS, respectively. The corresponding symmetry relations for the bosonic propagators read
\begin{eqnarray}\label{eq28}
\begin{split}
&P({\rm{or}} \; C,D)_{o_1 o_2 m,o_3 o_4 n} ({\bf{q}}) \\
&{\kern 20pt}=[P({\rm{or}} \; C,D)_{o_3 o_4 n,o_1 o_2 m} ({\bf{q}})]^*
\end{split}
\end{eqnarray}
for PHS and
\begin{eqnarray}\label{eq29}
\begin{split}
&P_{o_1 o_2 m,o_3 o_4 n} ({\bf{q}}) = e^{i{\bf{q}} \cdot ({\bf{R}}_n  - {\bf{R}}_m )}\\
&{\kern 20pt}\times P_{o_2 ,o_1 , - {\bf{R}}_m ;o_4 ,o_3 , - {\bf{R}}_n } ({\bf{q}}),\\
&C({\rm{or}} \; D)_{o_1 o_2 m,o_3 o_4 n} ({\bf{q}}) = e^{i{\bf{q}} \cdot ({\bf{R}}_n  - {\bf{R}}_m )}\\
&{\kern 20pt}\times C({\rm{or}} \; D)_{o_4 ,o_3 , - {\bf{R}}_n ;o_2 ,o_1 , - {\bf{R}}_m } ( - {\bf{q}})
\end{split}
\end{eqnarray}
for RAS.

\subsection{Estimation of order parameters}\label{sec3B}
There are several methods to determine the leading instabilities and the momentum dependence of corresponding order parameters. In early FRG analyses, it was a common implementation to assume the initial form of the order parameters and trace the RG flow of the order parameters and susceptibilities \cite{ref23, ref27, ref28, ref29, ref30, ref31}. The renormalized order parameters depend on their initial choice of the momentum dependence, and therefore, these studies do not present an unbiased tool to determine the form factor of the order parameters. It makes the problem cumbersome to postulate all possibilities of the initial form of the order parameters, especially for the case of multiband models. Thus, in later FRG studies addressing multiband models, the effective interaction in the particular ordering channel was decomposed into different eigenmode contributions, and the eigenfunction corresponding to the most diverging eigenvalue was estimated to be the form factor of the order parameter in the channel \cite{ref24, ref26, ref32, ref33}. We identify the form factor of order parameters by considering the linear response of the system to virtual infinitesimal external fields coupled to the fermion bilinears \cite{ref36}.

For the singlet and triplet pairings, we can add the following Hamiltonians in momentum space, respectively,
\begin{eqnarray}\label{eq30}
\begin{split}
&H_{{\rm{sSC}}}  =  - \frac{\lambda }{2}[\sum\limits_{o,o',{\bf{k}},\sigma } {S_{oo'}^* ({\bf{k}},{\bf{q}} - {\bf{k}}) } \\
&{\kern 40pt}\times \sigma {\kern 1pt} c_{{\bf{k}},o,\sigma }^\dag  c_{{\bf{q}} - {\bf{k}},o', - \sigma }^\dag   + {\rm{H.c.}}],
\end{split}
\end{eqnarray}
\begin{eqnarray}
\begin{split}
&H_{{\rm{tSC}}}  =  - \frac{\lambda }{2}[\sum\limits_{o,o',{\bf{k}},\sigma } {T_{oo'}^* ({\bf{k}},{\bf{q}} - {\bf{k}})} \\
&{\kern 40pt}\times c_{{\bf{k}},o,\sigma }^\dag  c_{{\bf{q}} - {\bf{k}},o', - \sigma }^\dag   + {\rm{H.c.}}],
\end{split}
\end{eqnarray}
in which the coupling constants $S_{oo'} ({\bf{k}},{\bf{q}} - {\bf{k}})$ and $T_{oo'} ({\bf{k}},{\bf{q}} - {\bf{k}})$ have the following relations:
\begin{eqnarray}\label{eq31}
\begin{split}
&S_{oo'} ({\bf{k}},{\bf{q}} - {\bf{k}}) = S_{o'o} ({\bf{q}} - {\bf{k}},{\bf{k}}),\\
&{\rm{  }}T_{oo'} ({\bf{k}},{\bf{q}} - {\bf{k}}) =  - T_{o'o} ({\bf{q}} - {\bf{k}},{\bf{k}}).
\end{split}
\end{eqnarray}
Now, we briefly describe how the form factor of the order parameter in a singlet paring channel can be estimated.

\begin{figure}[h!]
	\begin{center}
		\includegraphics[width=8.6cm]{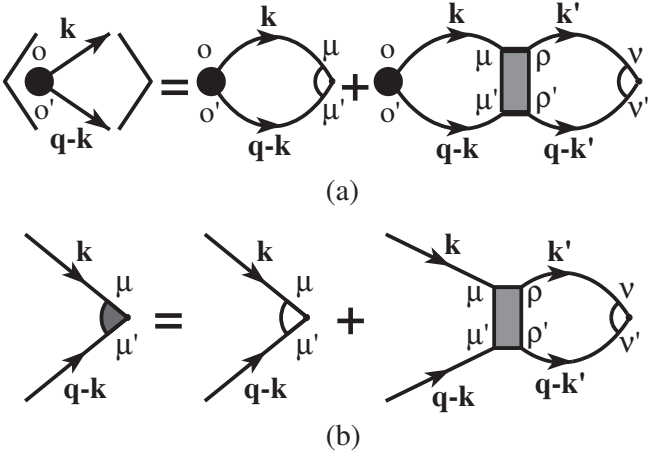}
	\end{center}
	\caption{(a) A schematic representation of the order parameter $\Delta _{oo'} ({\bf{k}},{\bf{q}} - {\bf{k}})$. The left diagram in the brackets stands for the order parameter. (b) The renormalization of the vertex $S_{\mu \mu '} ({\bf{k}},{\bf{q}} - {\bf{k}})$. The vertex with the gray arc is the renormalized three-point vertex, while that with the white arc is the bare vertex $S_{\mu \mu '} ({\bf{k}},{\bf{q}} - {\bf{k}})$.}
	\label{fig2}
\end{figure}

The additional action corresponding to Eq. (\ref{eq30}) is
\begin{eqnarray}\label{eq32}
\begin{split}
S_\lambda   = &  - \frac{\lambda }{2}[\sum\limits_{o,o',l,\sigma } {C_{o,o',l,\sigma }^* } \bar \psi _\sigma  (l,o)\bar \psi _{ - \sigma } (q - l,o')\\
& + \sum\limits_{o,o',l,\sigma } {C_{o,o',l,\sigma } } \psi _{ - \sigma } (q - l,o')\psi _\sigma  (l,o)],
\end{split}
\end{eqnarray}
where the index $l$ contains the wave vector $\bf{k}$ and Matsubara frequency $\omega$, i.e., $l = (\omega ,{\bf{k}})$, and the coupling constant $C_{o,o',l,\sigma}$ is $C_{o,o',l,\sigma }  = \sigma {\kern 1pt} S_{oo'} ({\bf{k}},{\bf{q}} - {\bf{k}})$.\\
The order parameter in the singlet paring channel is
\begin{eqnarray}\label{eq34}
\begin{split}
& \Delta_{oo'}^{\rm{sSC}} ({\bf{k}},{\bf{q}}-{\bf{k}}) \equiv \mathop{\lim} \limits_{\lambda\to+0}\sum\limits_\sigma \sigma \left\langle {c_{{\bf{k}},o,\sigma }^\dag c_{{\bf{q}}-{\bf{k}},o',-\sigma }^\dag}\right\rangle_\lambda,  
\end{split}
\end{eqnarray}
where $\left\langle \cdot \right\rangle _\lambda$ means the average in the grand-canonical ensemble \cite{ref37} with the action including the additional term in Eq. (\ref{eq32}),
\begin{eqnarray}\label{eq35}
\begin{split}
&\left\langle O \right\rangle _\lambda   = \frac{{\int {\prod\limits_{\xi ,\sigma } {d\bar \psi _\sigma  (\xi )} d\psi _\sigma  (\xi ){\kern 1pt} O} {\kern 1pt} e^{ - [S(\bar \psi ,\psi ) + S_\lambda  (\bar \psi ,\psi )]} }}{{\int {\prod\limits_{\xi ,\sigma } {d\bar \psi _\sigma  (\xi )} d\psi _\sigma  (\xi )} {\kern 1pt} e^{ - [S(\bar \psi ,\psi ) + S_\lambda  (\bar \psi ,\psi )]} }}.
\end{split}
\end{eqnarray}
When the system has no long-range order, the order parameter $\Delta_{oo'}^{\rm{sSC}} ({\bf{k}},{\bf{q}} - {\bf{k}})$ in Eq. (\ref{eq34}) vanishes. However, when the system is approaching the critical point, the corresponding susceptibility diverges, and the order parameter can develop. We postulate that the system has no long-range order and consider the quantity
\begin{eqnarray}
\begin{split}
\Delta _{oo'}^\lambda  ({\bf{k}},{\bf{q}} - {\bf{k}}) \equiv \sum\limits_\sigma  \sigma  \left\langle {c_{{\bf{k}},o,\sigma }^\dag  c_{{\bf{q}} - {\bf{k}},o', - \sigma }^\dag  } \right\rangle _\lambda \\  
= \frac{1}{\beta }\sum\limits_{\sigma ,\omega } \sigma  \left\langle {\bar \psi _\sigma  (l,o)\bar \psi _{ - \sigma } (q - l,o')} \right\rangle _\lambda . 
\end{split}\nonumber
\end{eqnarray}
Taking into account the vanishing of $\Delta _{oo'}^\lambda  ({\bf{k}},{\bf{q}} - {\bf{k}})$ at $\lambda  = 0$ and Taylor expanding Eq. (\ref{eq35}) with respect to $\lambda$, we have
\begin{eqnarray}\label{eq36}
\begin{split}
& \Delta _{oo'}^\lambda ({\bf{k}},{\bf{q}} - {\bf{k}}) = \frac{\lambda }{\beta }\sum\limits_{\sigma ,\omega } {\frac{\sigma }{2}} \sum\limits_{\nu ,\nu ',l',\sigma '} {C_{\nu ,\nu ',l',\sigma '} }  \\
& \kern 15pt \times \left\langle {\bar \psi _\sigma  (l,o)\bar \psi _{ - \sigma } (q - l,o')} \right.\left. {\psi _{ - \sigma '} (q - l',\nu ')\psi _{\sigma '} (l',\nu )} \right\rangle _{\lambda  = 0} .
\nonumber
\end{split}
\end{eqnarray}
Using the relation between the four-point Green's function and the effective interaction \cite{ref37} and taking an approximation of $G_{oo'} (\omega ,{\bf{k}}) \approx G_{oo'}^0 (\omega ,{\bf{k}})$, we obtain the following result:
\begin{eqnarray}\label{eq37}
\begin{split}
\Delta _{oo'}^\lambda ({\bf{k}},{\bf{q}} - {\bf{k}}) = 2\lambda &\sum\limits_{\mu ,\mu '} {L_{\mu o,\mu 'o'} ({\bf{k}},{\bf{q}} - {\bf{k}})} S_{\mu \mu '} ({\bf{k}},{\bf{q}} - {\bf{k}}) \\ 
- 2\lambda\sum\limits_{\nu ,\nu '} \sum\limits_{\rho ,\rho '} \sum\limits_{\mu ,\mu '}& \frac{1}{S_{BZ} } \int {d{\bf{k'}}} S_{\nu \nu '} ({\bf{k'}},{\bf{q}} - {\bf{k'}}) \\
\times L_{\nu \rho ,\nu '\rho '} (&{\bf{k'}},{\bf{q}}-{\bf{k'}}) L_{\mu o,\mu 'o'} ({\bf{k}},{\bf{q}} - {\bf{k}}) \\
\times& V_{\rho '\rho ,\mu '\mu } ({\bf{q}} - {\bf{k'}},{\bf{k'}};{\bf{q}} - {\bf{k}},{\bf{k}}),
\end{split}
\end{eqnarray}
where $V_{o'_1 o'_2 ,o_1 o_2 } ({\bf{k}}'_1 ,{\bf{k}}'_2 ;{\bf{k}}_1 ,{\bf{k}}_2 )$ is the effective interaction,
\begin{eqnarray}
\nonumber V_{o'_1 o'_2 ,o_1 o_2 } ({\bf{k}}'_1 ,{\bf{k}}'_2 ;{\bf{k}}_1 ,{\bf{k}}_2 ) = \mathop {\lim }\limits_{\Omega  \to 0} V_{o'_1 o'_2 ,o_1 o_2 }^\Omega  ({\bf{k}}'_1 ,{\bf{k}}'_2 ;{\bf{k}}_1 ,{\bf{k}}_2 ),
\end{eqnarray} 
and $L_{oo',\mu \mu '} ({\bf{k}},{\bf{q}} - {\bf{k}})$ is defined as
\begin{eqnarray}
L_{oo',\mu \mu'} ({\bf{k}},{\bf{q}} - {\bf{k}}) \equiv 
\frac{1}{\beta }\sum\limits_\omega{G_{oo'}^0 }
(\omega ,{\bf{k}})G_{\mu \mu '}^0 ( - \omega ,{\bf{q}} - {\bf{k}}).
\nonumber
\end{eqnarray}
Equation (\ref{eq37}) can be represented schematically as in Fig. \ref{fig2}(a) and is equivalent to the renormalization of the vertex $S_{\mu \mu '} ({\bf{k}},{\bf{q}} - {\bf{k}})$ shown in Fig. \ref{fig2}(b). The renormalization in Fig. \ref{fig2}(b) is consistent with Fig. 1 in Ref. \onlinecite{ref38}, where the form factors of order parameters were analyzed using the Bethe-Salpeter equation.

In the limit of $\lambda  \to  + 0$, the first term on the right-hand side of Eq. (\ref{eq37}) vanishes, while the second term may have a finite value due to a divergence of the effective interaction. So replacing $\bf{k}$ with ${\bf{k}} + {\bf{q}}$ in Eq. (\ref{eq37}), we get
\begin{eqnarray} \label{eq39}
\begin{split}
& \mathop {\lim }\limits_{\lambda  \to  + 0} \Delta _{oo'}^\lambda  ({\bf{k}} + {\bf{q}}, - {\bf{k}}) =  - 2 \mathop {\lim }\limits_{\lambda  \to  + 0} \lambda \sum\limits_{\nu ,\nu '} \sum\limits_{\rho ,\rho '} \sum\limits_{\mu ,\mu '} \\
& \times \frac{1}{{S_{BZ} }} \int {d{\bf{k'}}} S_{\nu \nu '} ({\bf{k'}} + {\bf{q}}, - {\bf{k'}}) L_{\nu \rho ,\nu '\rho '} ({\bf{k'}} + {\bf{q}}, - {\bf{k'}}) {\kern 20pt}\\ 
& \times V_{\rho \rho ',\mu \mu '} ({\bf{k'}} + {\bf{q}}, - {\bf{k'}};{\bf{k}} + {\bf{q}}, - {\bf{k}})L_{\mu o,\mu 'o'} ({\bf{k}} + {\bf{q}}, - {\bf{k}}). 
\end{split}
\end{eqnarray}
We can rewrite $V_{\rho \rho ',\mu \mu '} ({\bf{k'}} + {\bf{q}}, - {\bf{k'}};{\bf{k}} + {\bf{q}}, - {\bf{k}})$ in terms of $V_{\rho \rho 'm,\mu \mu 'n}^{\rm P} ({\bf{q}})$ by inverting Eq. (\ref{eq12}). 

By Fourier transforming Eq. (\ref{eq34}), we can get the order parameter in real space:
\begin{eqnarray}
\begin{split}
\Pi&_{oo'}^{\rm{sSC}}({\bf{R}}_i, {\bf{R}}_i-{\bf{R}}_\alpha) \\
&\equiv \mathop {\lim }\limits_{\lambda \to +0} \sum\limits_\sigma  \sigma
	\left\langle {c_{{\bf{R}}_i ,o,\sigma }^\dag
	c_{{\bf{R}}_i-{\bf{R}}_\alpha  ,o', - \sigma }^\dag } \right\rangle _\lambda   \\
&=\sum\limits_{\bf{q}} {e^{-i{\bf{q}}\cdot{\bf{R}}_i}} 
	\mathop {\lim }\limits_{\lambda \to +0}	\frac{1}{{S_{BZ} }}
	\int {d{\bf{k}}}f_\alpha^*({\bf{k}})\Delta_{oo'}^\lambda
	({\bf{k}} + {\bf{q}},-{\bf{k}}).
\end{split}\nonumber
\end{eqnarray}
Using Eq. (\ref{eq39}) and introducing $S_{oo'm} ({\bf{q}})$ by $S_{oo'} ({\bf{k}} + {\bf{q}}, - {\bf{k}}) = \sum\limits_m {S_{oo'm}^* ({\bf{q}})} f_m ({\bf{k}})$, we obtain the following expression for the order parameter:
\begin{eqnarray}\label{eq40}
\begin{split}
& \Pi_{oo'}^{\rm{sSC}} ({\bf{R}}_i , {\bf{R}}_i - {\bf{R}}_\alpha  ) =
	2 \mathop {\lim }\limits_{\lambda  \to  + 0} \lambda
	\sum \limits_{\bf{q}} {e^{-i{\bf{q}}\cdot{\bf{R}}_i} } \\
& {\kern 5pt} \times \sum\limits_{\mu ,\mu ',m} {S_{\mu \mu 'm}^* } ({\bf{q}}) \cdot
	\{\chi ^{{\rm{pp}}} ({\bf{q}})[-V^{\rm P} ({\bf{q}})]\chi ^{\rm{pp}} ({\bf{q}})\}_{\mu\mu'm,oo'\alpha} ,
\end{split}
\end{eqnarray}
where $\chi ^{{\rm{pp}}} ({\bf{q}})$ is
\begin{eqnarray}\label{eq41}
\begin{split}
&\chi _{o'_1 o'_2 m,o_1 o_2 n}^{{\rm{pp}}} ({\bf{q}}) = - \frac{1}{{S_{BZ} }}\int {d{\bf{k}}} f_m ({\bf{k}})f_n^* ({\bf{k}})\\
&{\kern 10pt}\times[\frac{1}{\beta }\sum\limits_\omega  {G_{o'_1 o_1 }^0 } (\omega ,{\bf{k}} + {\bf{q}})G_{o'_2 o_2 }^0 ( - \omega , - {\bf{k}})].
\end{split}
\end{eqnarray}
If only one pairing mode emerges with the transfer momentum $\bf{Q}$, the sum $\sum_{\bf{q}} {}$ in Eq. (\ref{eq40}) is removed, and $\bf{q}$ is replaced with $\bf{Q}$. We can use the eigenvalues $\Lambda _\beta  ({\bf{Q}})$ and eigenvectors $\phi _{o'_1 o'_2 m}^\beta  ({\bf{Q}})$ to decompose the following Hermitian matrix:
\begin{eqnarray}\label{eq42}
\nonumber & W_{o'_1 o'_2 m,o_1 o_2 n}^{{\rm{SC}}} ({\bf{Q}}) \equiv \{\chi ^{{\rm{pp}}} ({\bf{Q}})[-V^{\rm P} ({\bf{Q}})]\chi ^{{\rm{pp}}} ({\bf{Q}}) \}_{o'_1 o'_2 m,o_1 o_2 n} \\
& = \sum\limits_\beta  {\Lambda _\beta  ({\bf{Q}})} {\kern 1pt} {\kern 1pt} \phi _{o'_1 o'_2 m}^\beta  ({\bf{Q}})[\phi _{o_1 o_2 n}^\beta  ({\bf{Q}})]^* .
\end{eqnarray}
If only one eigenvalue $\Lambda _1 ({\bf{Q}})$ is dominantly divergent in FRG flow, we can expect the superconducting long-range order with the form factor represented by the corresponding eigenvector. In this case, Eq. (\ref{eq40}) becomes
\begin{eqnarray}
\begin{split}
\Pi&_{oo'}^{\rm{sSC}} ({\bf{R}}_i ,{\bf{R}}_i  - {\bf{R}}_\alpha  ) = 2e^{ - i{\bf{Q}} \cdot {\bf{R}}_i } \\
&\times \mathop {\lim }\limits_{\lambda  \to  + 0} \lambda \Lambda _1 ({\bf{Q}})[\sum\limits_{\mu ,\mu ',m} {S_{\mu \mu 'm}^* } ({\bf{Q}})\phi _{\mu \mu 'm}^1 ({\bf{Q}})] \cdot [\phi _{oo'\alpha }^1 ({\bf{Q}})]^*,
\end{split}\nonumber
\end{eqnarray}
which means
\begin{eqnarray}\label{eq43}
\begin{split}
\Pi _{oo'}^{\rm{sSC}} ({\bf{R}}_i ,{\bf{R}}_i  - {\bf{R}}_\alpha  ) = C e^{ - i{\bf{Q}} \cdot {\bf{R}}_i } [\phi _{oo'\alpha }^1 ({\bf{Q}})]^*. 
\end{split}
\end{eqnarray}
When several eigenvalues are similarly dominant in FRG flow, the system may have no long-range order due to competing effects between the different ordering tendencies. For the case of triplet pairing, all the results above are valid except for different symmetry relations of eigenvectors:
\begin{eqnarray}\label{eq44}
\begin{split}
[\phi _{oo'\alpha }^\beta  ({\bf{Q}})]^*  = & [\phi _{o'o\bar \alpha }^\beta  ({\bf{Q}})]^* e^{i{\bf{Q}} \cdot {\bf{R}}_\alpha  },\\
[\phi _{oo'\alpha }^\beta  ({\bf{Q}})]^*  = & - [\phi _{o'o\bar \alpha }^\beta  ({\bf{Q}})]^* e^{i{\bf{Q}} \cdot {\bf{R}}_\alpha  }\\
\end{split}
\end{eqnarray}
for singlet and triplet pairing, respectively, where $\bar \alpha$ is the index of the plane-wave basis with vector $-{\bf{R}}_\alpha$.

The form factors of order parameters in spin and charge channels can be obtained in a similar way and have nearly identical structures. The matrix $W^{{\rm{SC}}} ({\bf{Q}})$ in Eq. (\ref{eq42}) is changed into the matrices $W^{{\rm{SPN}}} ({\bf{Q}})$ for the spin channel and $W^{{\rm{CHG}}} ({\bf{Q}})$ for the charge channel,
\begin{eqnarray}\label{eq45}
& W^{{\rm{SPN}}} ({\bf{Q}}) = \chi ^{{\rm{ph}}} ({\bf{Q}})V^{\rm C} ({\bf{Q}})\chi ^{{\rm{ph}}} ({\bf{Q}}),\\
\nonumber & W^{{\rm{CHG}}} ({\bf{Q}}) = \chi ^{{\rm{ph}}} ({\bf{Q}})[V^{\rm C} ({\bf{Q}}) - 2V^{\rm D} ({\bf{Q}})]\chi ^{{\rm{ph}}} ({\bf{Q}}).
\end{eqnarray}
Here  $\chi ^{{\rm{ph}}} ({\bf{q}})$ is
\begin{eqnarray}\label{eq46}
\begin{split}
&\chi _{o'_1 o'_2 m,o_1 o_2 n}^{{\rm{ph}}} ({\bf{q}}) =  - \frac{1}{{S_{BZ} }}\int {d{\bf{k}}} f_m ({\bf{k}})f_n^* ({\bf{k}})\\
&{\kern 10pt}\times[\frac{1}{\beta }\sum\limits_\omega  {G_{o'_1 o_1 }^0 } (\omega ,{\bf{k}} + {\bf{q}})G_{o_2 o'_2 }^0 (\omega ,{\bf{k}})].
\end{split}
\end{eqnarray}
The order parameters in real space are defined as
\begin{eqnarray}\label{eq47}
\nonumber \Pi _{oo'}^{\rm{tSC}} ({\bf{R}}_i ,{\bf{R}}_i  - {\bf{R}}_\alpha  ) = 
	\mathop {\lim }\limits_{\lambda  \to  + 0} \sum\limits_\sigma 
	{\left\langle {c_{{\bf{R}}_i ,o,\sigma }^\dag 
	c_{{\bf{R}}_i  - {\bf{R}}_\alpha  ,o', - \sigma }^\dag  } \right\rangle _\lambda  },\\
\nonumber \Pi _{oo'}^{\rm{SPN}} ({\bf{R}}_i ,{\bf{R}}_i  - {\bf{R}}_\alpha  ) =
	\mathop {\lim }\limits_{\lambda  \to  + 0} \sum\limits_\sigma  \sigma
	\left\langle {c_{{\bf{R}}_i ,o,\sigma }^\dag
	c_{{\bf{R}}_i  - {\bf{R}}_\alpha  ,o',\sigma } } \right\rangle _\lambda, \\
\nonumber \Pi _{oo'}^{\rm{CHG}} ({\bf{R}}_i ,{\bf{R}}_i  - {\bf{R}}_\alpha  ) =
	\mathop {\lim }\limits_{\lambda  \to  + 0} \sum\limits_\sigma
	{\left\langle {c_{{\bf{R}}_i ,o,\sigma }^\dag
	c_{{\bf{R}}_i  - {\bf{R}}_\alpha  ,o',\sigma } } \right\rangle _\lambda  } {\kern 6pt}\\
\end{eqnarray}
for the triplet pairing channel, the spin channel, and the charge channel, respectively. Similar to Eq. (\ref{eq43}), the form factor in the spin (charge) channel is given by
\begin{eqnarray}\label{eq48}
\begin{split}
& \Pi _{oo'}^{\rm{SPN (CHG)}} ({\bf{R}}_i ,{\bf{R}}_i  - {\bf{R}}_\alpha  ) \\
& {\kern 20pt} = C{\kern 1pt} e^{ - i{\bf{Q}} \cdot {\bf{R}}_i }
	[\phi _{oo'\alpha }^1 ({\bf{Q}})]^* 
+ C^* e^{i{\bf{Q}}\cdot({\bf{R}}_i-{\bf{R}}_\alpha)}
	\phi_{o'o\bar \alpha }^1({\bf{Q}}),{\kern 10pt}
\end{split}
\end{eqnarray}
where $\phi _{oo'\alpha }^1 ({\bf{Q}})$ is the eigenvector of the dominantly divergent eigenmode of the matrix $W^{{\rm{SPN}}}({\bf{Q}})$ [$W^{{\rm{CHG}}}({\bf{Q}})$] in the spin (charge) channel.

\section{Results and Discussion}\label{sec4}

Fourier transforming and diagonalizing the single-particle Hamiltonian, Eq. (\ref{eq02}), we obtain the one-particle energy and transformation matrix in Eq. (\ref{eq13}),
\begin{eqnarray}\label{eq49}
\begin{split}
 E_1( & {\bf{k}}) =  - |d({\bf{k}})|, E_2 ({\bf{k}}) =  + |d({\bf{k}})|,\\
& T_{ob}({\bf{k}})=\left(\begin{array}{cc} \frac{d({\bf{k}})}{|d({\bf{k}})|}
\quad \frac{d({\bf{k}})}{|d({\bf{k}})|}\\ -1 {\kern 23pt} 1{\kern 6pt} \end{array}\right),
\end{split}
\end{eqnarray}
with $d({\bf{k}}) = t(1+2\cos\frac{{k_x }}{2}e^{-i\frac{{\sqrt 3 }}{2}k_y })$. The initial values of projections, $V^{P,(0)}({\bf{q}})$, $V^{\rm C,(0)}({\bf{q}})$, and $V^{\rm D,(0)}({\bf{q}})$, are obtained by Fourier transforming the interaction Hamiltonian in Eq. (\ref{eq03}) and projecting it onto the three channels via Eq. (\ref{eq12}). The expressions of these values are very simple in the orbital picture:
\begin{eqnarray}\label{eq51}
\begin{split}
& V_{AA0,AA0}^{\rm P(C),(0)} ({\bf{q}}) = V_{BB0,BB0}^{\rm P(C),(0)} ({\bf{q}}) = U,\\ 
& V_{AA0,AA0}^{\rm D,(0)} ({\bf{q}}) = V_{BB0,BB0}^{\rm D,(0)} ({\bf{q}}) = U + V_2 \sum\limits_{j = 1}^6 {e^{i{\bf{q}} \cdot {\bf{R}}_j } },\\
& V_{AAm,AAm}^{\rm P(C),(0)} ({\bf{q}}) = V_{BBm,BBm}^{\rm P(C),(0)} ({\bf{q}}) = V_2 \;\; (m = 1 \sim 6),\\
& V_{AB0,BA0}^{\rm P(C,D),(0)} ({\bf{q}}) = V_{BA0,AB0}^{\rm P(C,D),(0)} ({\bf{q}}) = J,\\
& V_{ABm,BA\bar m}^{\rm P(C,D),(0)} ({\bf{q}}) = [V_{BA\bar m,ABm}^{\rm P(C,D),(0)} ({\bf{q}})]^ * \\
&       \kern 75pt = Je^{ - i{\bf{q}} \cdot {\bf{R}}_m } \;\; (m = 2,3),\\
& V_{AA0,BB0}^{\rm P(C),(0)} ({\bf{q}}) = [V_{BB0,AA0}^{\rm P(C),(0)} ({\bf{q}})]^*  \\
&       \kern 75pt = J(1 + e^{ - i{\bf{q}} \cdot {\bf{R}}_2 }  + e^{ - i{\bf{q}} \cdot {\bf{R}}_3 } ),\\
& V_{ABm,ABm}^{\rm D,(0)} ({\bf{q}}) = J \;\; (m = 0,2,3),\\
& V_{BAm,BAm}^{\rm D,(0)} ({\bf{q}}) = J \;\; (m = 0,5,6),\\
& {\rm{All \; other \; elements}} = 0.\\
\end{split}
\end{eqnarray}
\begin{figure}[h!]
	\begin{center}
		\includegraphics[width=8.6cm]{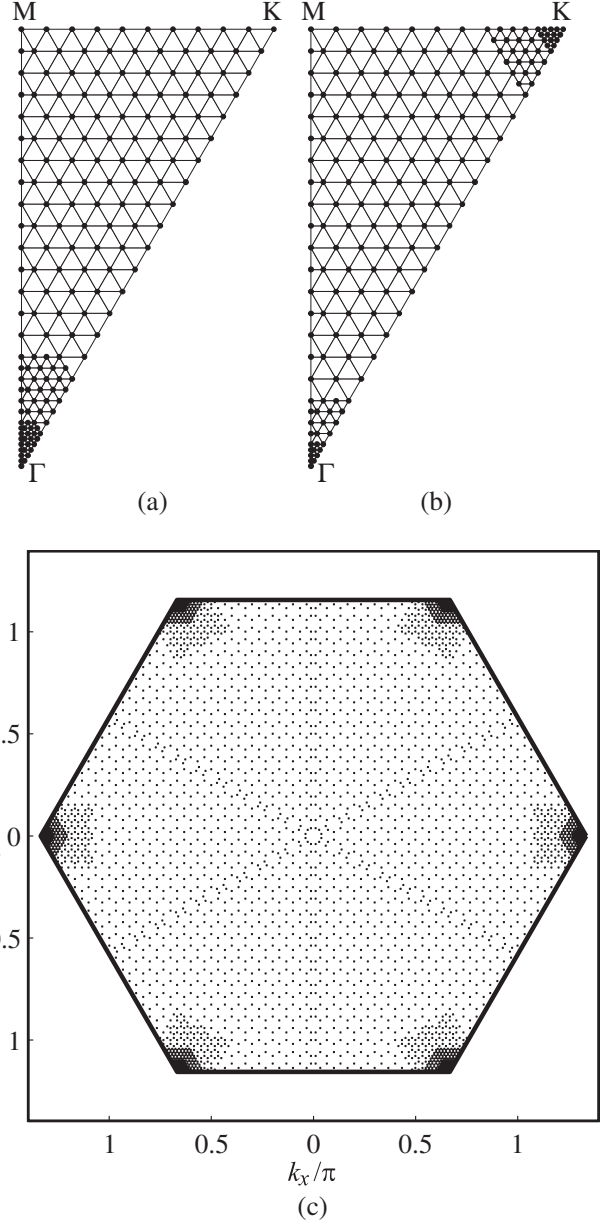}
	\end{center}
	\caption{(a) Mesh of $N_q  = 171$ points for transfer momenta within the irreducible region of the BZ in the particle-particle channel. The points are distributed more densely near the ${\bf{\Gamma }}$ point. The bosonic propagators $P(\bf{q})$ are calculated for these points. (b) Mesh of $N_q  = 175$ points for transfer momenta within the irreducible region of the BZ in the particle-hole channel. The points are distributed more densely near the ${\bf{\Gamma }}$ and $\bf{K}$ points. The bosonic propagators $C(\bf{q})$ and $D(\bf{q})$ are calculated for these points. (c) Mesh of $N_k  = 8280$ points for sampling momenta used in the integration of $\dot \chi ^{{\rm{pp}}}$ and $\dot \chi ^{{\rm{ph}}}$. The points are distributed more densely near the Dirac points $\bf{K}$ and ${\bf{K'}}$.}
	\label{fig3}
\end{figure}

In the current implementation of TUFRG, the matrices $P$, $C$, $D$, $V^{\rm P}$, $V^{\rm C}$, $V^{\rm D}$, $\chi^{{\rm{pp}}}$, and $\chi^{{\rm{ph}}}$ have $52\times52$ structures. The flow equation for the bosonic propagators, Eq. (\ref{eq10}), is solved for the transfer momenta in the irreducible region of the BZ shown in Figs. \ref{fig3}(a) and \ref{fig3}(b). The discretized transfer momenta are distributed more densely where the ordering vectors are expected. Figure \ref{fig3}(c) shows the sampling momenta used for the integration of $\dot \chi ^{{\rm{pp}}}$ and $\dot \chi ^{{\rm{ph}}}$ in Eq. (\ref{eq11}), which are denser near the Dirac points ${\bf{K}}$ and ${\bf{K'}}$. This ${\bf{k}}$ mesh is obtained in way similar to that in Ref. \onlinecite{ref26}.

\begin{figure}[h!]
	\begin{center}
		\includegraphics[width=8.6cm]{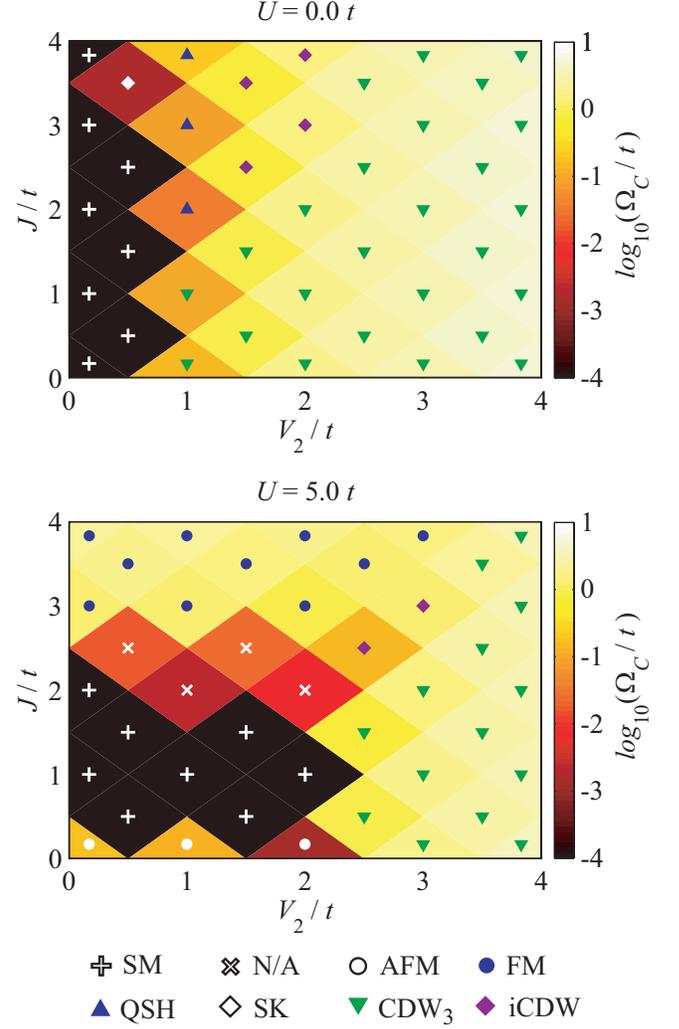}
	\end{center}
	\caption{(Color online) Dominant instabilities for different interaction parameters. The color bar indicates the value of the critical scales $\Omega _C$ at which the corresponding transitions may occur. In the region marked SM, there is no divergence of any bosonic propagator in the RG flow down to the stopping scale $\Omega ^*=10^{-4}t$. In the region marked N/A, there is a divergence of some bosonic propagator; however, different ordering tendencies coexist and compete with each other, so that a clear identification of the leading instability is not possible, and the divergent $W^{{\rm{SC}}} ({\bf{Q}}),W^{{\rm{SPN}}} ({\bf{Q}})$, or $W^{{\rm{CHG}}} ({\bf{Q}})$ at $\Omega _C$ shows a mix of various instabilities.}
	\label{fig4}
\end{figure}

The different tendencies towards symmetry-broken ground states are identified via the estimation method in Sec. \ref{sec3B}. We have analyzed these tendencies by varying the interaction parameters $V_2$ and $J$ while fixing the on-site repulsion $U$. The results are summarized in the tentative phase diagrams shown in Fig. \ref{fig4}. The critical scales $\Omega _C$ at which these transitions may occur are also provided using the color bar. We explain the instabilities appearing in the phase diagram below.

{\it Antiferromagnetic spin-density-wave instability.}
It is known that the on-site repulsion exceeding a critical value $U_C \approx 3.8t$ drives the antiferromagnetic (AFM) spin-density-wave instability for the half-filled Hubbard model on the honeycomb lattice \cite{ref13}. This instability is manifested in the RG flow as a dominantly divergent eigenmode of $W^{{\rm{SPN}}}({\bf{Q}}=0)$ which has real numbers $\phi_{AA0}$ and $\phi_{BB0}$ with the relation $\phi_{AA0}=-\phi_{BB0}$ as its largest components. The relation $\phi_{AA0}=-\phi_{BB0}$ represents the staggered spin configuration on the $A$ and $B$ sublattices. The AFM instability occurs only for dominant $U$ ($U=5.0t$ in Fig. \ref{fig4}) and vanishes with the inclusion of small $J$. The vanishing of the AFM with small $J$ implies that the exchange interaction $J$ has a strong tendency to destroy AFM and recover the semimetallic (SM) phase. As can be seen in Fig. \ref{fig4}, the inclusion of parameter $J$ smaller than $1.5t$ has no impact on the emergent phases for $U=0$, which is supposed to be due to two competing ordering tendencies of $J$, i.e., the pairing and ferromagnetic tendencies. Involving $V_2$ larger than $2.5t$ also induces the vanishing of AFM, but it makes another charge-density-wave order develop. The SM phase is stable in an extended region, implying the suppression of spin or charge order due to competition effects between the different ordering tendencies \cite{ref14}.

{\it Ferromagnetic spin-density-wave instability.}
This instability is manifested in the RG flow as a dominantly divergent eigenmode of $W^{{\rm{SPN}}} ({\bf{Q}} = 0)$ which has real numbers $\phi _{AA0}  = \phi _{BB0}$ as its largest components. Increasing $J$ larger than $2.8t$ drives the ferromagnetic (FM) spin-density-wave instability for $U=5.0t$ but not for $U=0$. This fact can be explained by the duality of the effect of the exchange interaction $J$, the ferromagnetic exchange that promotes the spin alignment and the pair hopping that drives the superconducting pairing. When $U=5.0t$, the large value of the on-site repulsion $U$ blocks the pair hopping and permits only the ferromagnetic exchange. The ferromagnetic exchange ultimately wins the AFM ordering tendency by the parameter $U$ and produces the FM ordered phase when increasing $J$. However, for the case of zero on-site repulsion, the absence of the Coulomb blockade effect induces the competition between the ferromagnetic exchange and pair hopping, thus suppressing the FM order.

{\it Quantum spin hall instability.}
The QSH phase is the most tempting one in the phase diagram in Fig. \ref{fig4}. The existence of this phase remains an inconclusive problem. The QSH instability is characterized by the following dominantly divergent eigenmode of $W^{{\rm{SPN}}}({\bf{Q}} = 0)$:
\begin{eqnarray}\label{eq52}
\begin{split}
 \phi _{AA1}  =  - \phi _{AA2}  = \phi _{AA3}  =  - \phi _{AA4} \\
 = \phi _{AA5}  =  - \phi _{AA6}  =  - iR,\\
 \phi _{BB1}  =  - \phi _{BB2}  = \phi _{BB3}  =  - \phi _{BB4} \\
 = \phi _{BB5}  =  - \phi _{BB6}  = iR,
\end{split}
\end{eqnarray}
with a real constant $R$.
\begin{figure}[h!]
	\begin{center}
		\includegraphics[width=8.6cm]{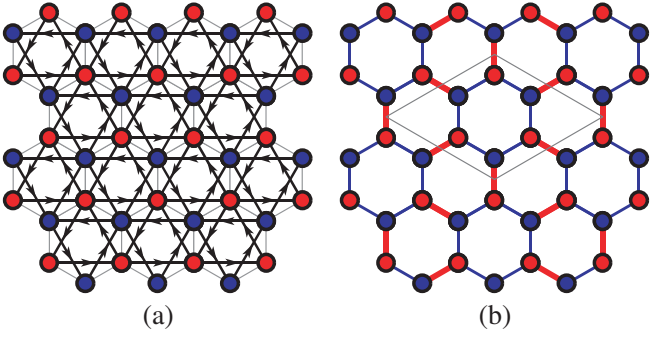}
	\end{center}
	\caption{(Color online) Representative spin current and bond strength patterns for the quantum spin Hall phase and spin-Kekul\'{e} phase. (a) Spin current pattern in the quantum spin Hall phase. The arrows indicate the directions of the spin currents. (b) Spin bond strength pattern in the spin-Kekul\'{e} phase. The red (blue) lines represent the positive (negative) values of spin bond order parameters, and the linewidths indicate their magnitudes. A unit cell is shown as the region enclosed by the gray line.}
	\label{fig5}
\end{figure}
This form factor corresponds to an ordered pattern of spin currents shown in Fig. \ref{fig5}(a). We have not found the QSH instability in the absence of the exchange interaction $J$, which is consistent with a previous TUFRG result \cite{ref14}. In our study it occurs in a relatively narrow region of the parameter space, namely, around $U \approx 0,V_2 \approx t,J = 2t \sim 4t$. We performed test calculations in which we involved only the pair hopping or the ferromagnetic exchange, separately, keeping the values of $J$ and $V_2$ unchanged. Involving only one effect did not drive the QSH order, which demonstrated that a combination of pair hopping and ferromagnetic exchange is essential for the emergence of QSH. Since the density-density repulsion $V_2$ is generally larger than the exchange integral $J$, the region of QSH phase is far from reality. A large value of $J$ can affect the validity of the weak-coupling FRG approach and damage the reliability of the result. However, we expect that our results could shed some light on the issue of the existence of a QSH state.

{\it Spin-Kekul\'{e} bond order instability.}
This instability is manifested in the RG flow as the following divergent eigenmode of $W^{{\rm{SPN}}} ({\bf{Q}} = {\bf{K}})$ in the spin channel:
\begin{eqnarray}\label{eq53}
\begin{split}
& \phi _{AB0}  = \phi _{BA0}  = R, \phi _{AB2}  = \phi _{BA5}^*  = R e^{ - i\frac{{2\pi }}{3}} , \\
&          \kern 45pt \phi _{AB3}  = \phi _{BA6}^*  = R e^{i\frac{{2\pi }}{3}}, 
\end{split}
\end{eqnarray}
which represents a spin bond order with an enlarged unit cell, as shown in Fig. \ref{fig5}(b). This order can be thought of as the spinful counterpart of the Kekul\'{e} bond order \cite{ref07, ref10}. The spin-Kekul\'{e} (SK) phase was proposed theoretically in a previous study \cite{ref04}. Its unit cell is three times bigger than the original one and contains six atoms. Although our current Hamiltonian is capable of producing this exotic spin-Kekul\'{e} phase, the fact that it emerges only in one segment of our phase diagram with dominant $J$ ($U \approx 0, V_2  \approx 0.5t, J \approx 3.5t$) casts some doubt on this result. We cannot exclude the possibility that the SK phase might be an artifact generated by the enhancement of the exchange integral.
\begin{figure}[h!]
	\begin{center}
		\includegraphics[width=8.6cm]{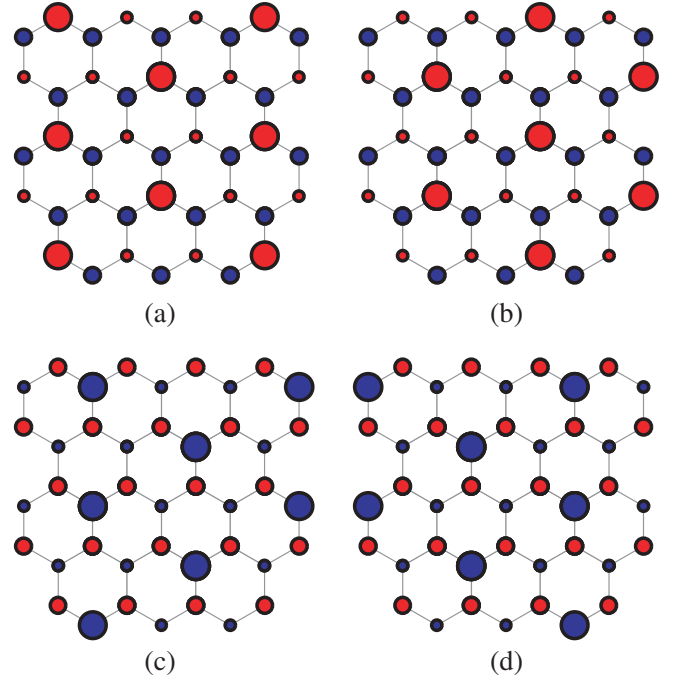}
	\end{center}
	\caption{(Color online) Representative charge ordering patterns for the three-sublattice charge-density-wave phase. The real charge-density distribution in the phase is expressed by linear combination of these four patterns. The radius is a measure of the local charge density on each site.}
	\label{fig6}
\end{figure}

{\it Three-sublattice charge-density-wave instability.}
A large part of our phase diagrams is occupied by a CDW phase called the three-sublattice charge density wave ($\rm{CDW}_3$) \cite{ref14}, which has an enlarged unit cell like the spin-Kekul\'{e} phase. Especially, for the case of $U=0$, the $\rm{CDW}_3$ phase develops in the major part of the phase diagram. It is driven by the next-nearest-neighbor repulsion $V_2$ exceeding a critical value and is characterized by the following two degenerate divergent eigenmodes of $W^{{\rm{CHG}}} ({\bf{Q}} = {\bf{K}})$ in the charge channel:
\begin{eqnarray}\label{eq54}
\begin{split}
& \phi _{AA0}^1  = R, \kern 8pt \phi _{BB0}^1  = 0, \\
& \phi _{AA0}^2  = 0, \kern 8pt \phi _{BB0}^2  = R.
\end{split}
\end{eqnarray}
These two form factors are represented by the charge ordering patterns shown in Fig. \ref{fig6} and these four patterns will be linearly superposed to construct the charge density distribution in real space. To determine their superposition in detail, one needs to perform the mean-field calculation or the analysis beyond the linear response, but these are beyond the scope of the present work.

The low-energy effective Hamiltonian becomes
\begin{eqnarray}\label{eq55}
\begin{split}
& H_{{\rm{CDW}}_{\rm{3}} }=-\frac{1}{N}\sum\limits_o V (N_{\bf{K}}^o N_{ - {\bf{K}}}^o  + N_{ - {\bf{K}}}^o N_{\bf{K}}^o )
\end{split}
\end{eqnarray}
where $N_{\bf{K}}^o = \sum {_{{\bf{k}},\sigma } } c_{{\bf{k}} + {\bf{K}},o,\sigma }^\dag  c_{{\bf{k}},o,\sigma }$, and $V$ is a positive constant. Unlike the effective Hamiltonian in Ref. \onlinecite{ref14}, Eq. (\ref{eq55}) has no coupling between different orbital indices. Even when properly involving the nearest-neighbor repulsion $V_1$, the form of the low-energy effective Hamiltonian in Eq. (\ref{eq55}) is retained. This disagreement can be attributed to the so-called {\em orbital makeup} in the band picture.

{\it Incommensurate charge-density-wave instability.}
When increasing the parameter $J$, the $\rm{CDW}_3$ phase changes to an incommensurate charge-density-wave (iCDW) phase. The ordering vector depends on the value of $V_2$ and $J$, wandering near the $\bf{K}$ point. The iCDW phase was first reported in Ref. \onlinecite{ref14}. It is manifested in the RG flow as the following divergent eigenmode of $W^{{\rm{CHG}}} ({\bf{Q}} = {\bf{Q}}_0)$ in the charge channel:
\begin{equation}
\phi _{AA0}^1  = R e^{i\Delta \varphi } , \kern 8pt \phi _{BB0}^1  =  - R e^{ - i\Delta \varphi} ,
\end{equation}
which gives the charge distribution,
\begin{eqnarray}\label{eq56}
\begin{split}
& N_{{\bf{R}}_i }^A  = N_0  + \Delta N \cos ({\bf{Q}}_0  \cdot {\bf{R}}_i  + \varphi _0  + \Delta \varphi ),\\
& N_{{\bf{R}}_i }^B  = N_0  - \Delta N \cos ({\bf{Q}}_0  \cdot {\bf{R}}_i  + \varphi _0  - \Delta \varphi ).
\end{split}
\end{eqnarray}
In the transition from the $\rm{CDW}_3$ to iCDW phase, the degeneracy will change from 2 to 1, and the ordering vector will also change gradually near the $\bf{K}$ point. It would be interesting to investigate the behavior of this transition in detail, which we leave for future work.

For comparison, we have also analyzed the TUFRG result of the effective interaction and identified the leading instabilities towards ordered states by using the method described in Ref. \onlinecite{ref26}. The obtained phase diagrams are nearly identical to that in Fig. 4, with the only difference being in one segment of the parameter space. More specifically, for the parameter region of $U=0, V_2=t, J=2t,$ our approach gives the QSH instability, while the previous approach gives the iCDW instability. Taking into account the fact that this region is located at the border between the QSH and iCDW phases, we suppose that this deviation is essentially negligible.

\section{Conclusion} \label{sec5}

In this work we have investigated the effect of enhanced exchange interaction on possible ground-state orderings of electrons on the honeycomb lattice at half filling. In order to calculate the effective interactions and analyze the ground states of the system, we have employed the TUFRG scheme with a high resolution of wave-vector dependences in the effective interaction. The ground-state phase diagrams in the space of next-nearest-neighbor repulsion and nearest-neighbor exchange integral were obtained for two typical values of on-site repulsion, namely, for $U=0$ and $U=5t$. Inclusion of the exchange interaction produces a phase diagram with diverse ordering tendencies, especially for vanishing on-site repulsion. The commensurate CDW order, i.e., the $\rm{CDW}_3$ order, developed in a wide region of the diagram, which is replaced by the incommensurate CDW order when increasing the exchange integral. The topological QSH state emerged in a relatively narrow region of the parameter space around $U\approx 0, V_2\approx t, J = 2t \sim 4t$. This phase is induced by a combination of the ferromagnetic exchange and pair hopping interactions. Through a test calculation, we verified that involving only the pair hopping or the ferromagnetic exchange would not produce the QSH phase. There exists another interesting phase named the spin-Kekul\'{e} phase in a very small part of the parameter space near $U \approx 0, V_2\approx 0.5t, J\approx 3.5t$, but it is suspected to be an artifact of the TUFRG calculation due to the very limited region of existence and too large value of the exchange integral. For the case of $U=5t$, there also exist antiferromagnetic and ferromagnetic phases for small and large values of the parameter $J$, respectively, but these phases change into $\rm{CDW}_3$ by increasing $V_2$. There is strong competition between different ordering tendencies, which renders the semimetallic phase stable in relatively wide regions of the phase diagrams. Any signal for a dominating pairing instability has not been found at half filling as in previous results.

In addition, we derived the symmetry relations of the bosonic propagators and proposed the linear-response-based approach for identifying the type of order. The former can efficiently reduce the computational effort for systems with a high geometrical symmetry, and the latter can reasonably and quickly determine the form factors of order parameters.

\acknowledgments{
We thank Kwang-Hyok Jong, Kum-Hyok Jong, and Chol-Jun Kang for useful discussions. S.J.O. is grateful to Prof. Guang-Shan Tian for giving excellent lectures on the physics of strongly correlated systems.
}

\bibliographystyle{apsrev}
\bibliography{manuscriptV2}

\end{document}